\documentclass[aps,twocolumn,amsfonts,amssymb]{revtex4}

\usepackage{epsfig}

\begin{document}

\title{Testing Quantum Dynamics using Signaling}


\author{Aditi Sen(De) and Ujjwal Sen}

\affiliation{
ICFO-Institut de Ci\`encies Fot\`oniques, Jordi Girona 29, Edifici Nexus II, E-08034 Barcelona, Spain \\
Institut f\"ur Theoretische Physik, Universit\"at Hannover, D-30167 Hannover,
Germany
}

\begin{abstract}

We consider a physical system in which the description of states and measurements 
follow the usual quantum mechanical rules. We also assume that the dynamics is linear, but may not be 
fully quantum (i.e unitary).   We show that in such a physical system, certain \emph{complementary} evolutions, 
namely cloning and deleting operations that give a better fidelity than quantum mechanically allowed ones, 
in one (inaccessible) region, 
lead to signaling to a far-apart 
(accessible) region. 
To show such signaling, one requires certain two-party quantum correlated 
states shared between the 
two regions.  Subsequent measurements  are performed only in the accessible part to detect such phenomenon.

\end{abstract}

\maketitle

\newtheorem{lemma}{Lemma}
\newtheorem{corollary}{Corollary}
\newtheorem{theorem}{Theorem}

The existence of quantum correlation in states shared between distant partners
has several important fundamental and practical impacts \cite{NC}. 
One can obtain violation of local realism by using states with quantum correlation 
\cite{EPR}.
On the other hand, one may use states with quantum correlation in nonclassical 
tasks like cryptography \cite{crypto}, dense coding \cite{dense}, teleportation \cite{tele},
etc.

In this paper, we show that quantum correlations can be used to check for or to provide 
bounds on possible non-quantum 
effects. 
Non-quantum effects have generally been divided into two categories: Ones which are non-quantum 
in the ``statics'' part of the theory, and ones which are so in the ``dynamics'' part (see e.g. \cite{Mielnik}).
We consider a physical system in which 
(i) \emph{the states} (\(\left|\psi\right\rangle\), \(\left|\phi\right\rangle\), etc.) 
\emph{are elements of a complex Hilbert space}, just as in 
quantum mechanics. And (ii) \emph{measurements are also assumed just as in quantum mechanics}. 
The duo is said to form the ``statics'' part of the theory. 
We further assume that (iii) \emph{the dynamics is linear}, i.e. 
\(\left|\psi\right\rangle \rightarrow \left|\psi{'}\right\rangle\) and 
\(\left|\phi\right\rangle \rightarrow \left|\phi{'}\right\rangle\) implies 
\(a \left|\psi\right\rangle + b \left|\phi\right\rangle \rightarrow a \left|\psi{'}\right\rangle
+ b \left|\phi{'}\right\rangle\), for complex \(a\) and \(b\).
Note that (i), (ii), and (iii),
by themselves, does not imply a quantum dynamics (i.e. the usual unitary dynamics).
For our purposes, it is important to note that  (i) and (ii)
 lead to  quantum correlation in states of separated parties.  
We 
show that 
in such a physical system, 
certain \emph{complementary} families of 
non-quantum evolutions give rise to signaling. 
This gives us an independent basis to believe in the quantum dynamics.

\begin{figure}[ht]
\begin{center}
\unitlength=0.3mm
\begin{picture}(-20,160)(0,0)

\put(10,110){\vector(1,1){35}}
\put(10,110){\vector(1,-1){35}}

\put(10,106){\line(0,1){8}}
\put(7,106){\line(0,1){8}}

\put(7,106){\line(1,0){3}}
\put(7,114){\line(1,0){3}}

\put(7,110){\vector(-1,1){35}}
\put(7,110){\vector(-1,-1){35}}

\put(-28,75){\vector(-1,-1){65}}

\put(-135,0){to ``environment'' (B)}

\put(-35,85){Polarizer}

\put(-105,108){Pump pulse}
\put(-65,150){Trigger}

\put(50,150){to Alice (A)}

\put(55,65){\line(-1,1){14}}
\put(55,65){\line(-1,-1){46}}
\put(8.5,19){\line(-1,1){46}}

\put(32,124){1}
\put(32,91){2}
\put(-3,94){4}
\put(-3,122){3}

\put(82,107){DM}

\put(58,62){M}

\put(4,8){M} 

\put(-69,63){HSM}

\thicklines
\put(-40,110){\vector(1,0){30}}
\put(-10,110){\line(1,0){17}}
\put(10,110){\line(1,0){10}}
\put(30,110){\vector(-1,0){10}}
\put(30,110){\vector(1,0){10}}
\put(40,110){\line(1,0){40}}

\put(80,105){\line(0,1){10}}

\put(56,60){\line(0,1){10}}

\put(3.5,18){\line(1,0){10}}

\put(-42.5,65){\line(1,0){10}}

\end{picture}
\end{center}
\caption{Schematic description  of the 
arrangement in the case of cloning.
The down conversion crystal is denoted as a box, and
delay mirror, mirrors, and half-silvered mirror are denoted respectively by DM, M, and HSM. 
See text for details.
} 
\label{expt_figure}
\end{figure}
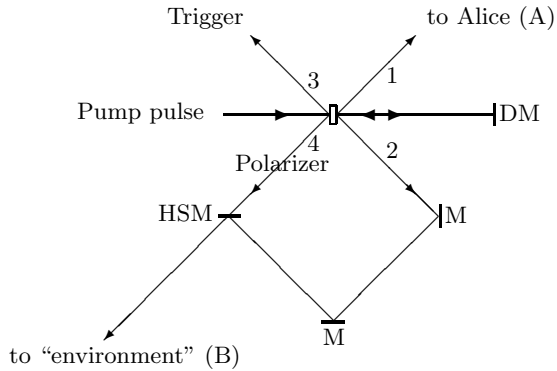

In checking for the effect, we will use cloning \cite{WZ} and deleting \cite{Braun_Pati} operations as our tools.
It was shown in \cite{MAU} that exact cloning or exact deleting, results in a change of 
von Neumann entropy \cite{von}. Within the quantum formalism,  although exact cloning and deleting are not 
possible, approximate versions of such operations \emph{are} possible (see e.g. \cite{Buzek, Braun_Pati2}). 
To check the effect, one 
requires to 
prepare certain bipartite states, which we show to be available within the reach of current technology.
Importantly,
we do not need to directly observe (perform measurements in) the region whose dynamics 
is being probed.
%
%
%
%
%
%
%
%
%
%
%
We suppose that one part (B) of the bipartite state is 
lost to the ``environment''.
The other
part (A) remains in the ``accessible'' part of the experiment (see Fig.
\ref{expt_figure}). 
In this paper 
we show, that in our physical system (i.e. one which follows (i), (ii), and (iii)), 
whenever the evolution in the 
\emph{environment} (B), 
is such that a cloning or deleting 
happens with a 
better 
fidelity
than the best quantum mechanical cloning or deleting machine, there occurs a change
of entropy in the \emph{accessible} part (A) of the experiment. 
This change of entropy can be detected in the A part, and therefore results in a 
signaling to the A part.  Note here that if we believe that signaling is not possible \cite{Shimony},
then our results prove that cloning and deleting (that are better than 
what can be done by the best quantum mechanical machines) are not possible, without assuming 
the whole quantum dynamics.
The reason for the choice of 
the two operations of cloning and deleting
is that it has been generally argued,
that they are in a sense complementary.
Thus, it is 
conceivable that at least one of such non-quantum mechanical operations
or ``nearby'' ones are possible to occur, \emph{if at all}, in 
the environment.

%
%
%
%

\emph{Cloning and deleting.} 
Let us first briefly consider the notions of cloning and deleting. In cloning, we want to have the 
evolution 
\(\left|\psi\right\rangle \left|0\right\rangle \rightarrow \left|\Psi\right\rangle, \quad 																											
\left|\phi\right\rangle \left|0\right\rangle \rightarrow \left|\Phi\right\rangle\),
where \(\left|0\right\rangle\) is a fixed  ``blank'' state in which the cloned state is to appear.
In the exact case, we want to have 
\(
\left|\Psi\right\rangle  =\left|\psi\right\rangle  \left|\psi\right\rangle\), 
 and \(\left|\Phi\right\rangle  = \left|\phi\right\rangle \left|\phi\right\rangle\).
This however is not possible under a quantum mechanical evolution, when 
\(\left|\psi \right\rangle\) and \(\left|\phi\right\rangle\) are not orthogonal 
\cite{WZ,Yuen}. Consequently, one 
may want to have the best cloning machine, i.e. one that takes \(\left|\Psi\right\rangle\) 
as close as possible to \(\left|\psi\right\rangle  \left|\psi\right\rangle\), and  
at the same time takes \(\left|\Phi\right\rangle\) as close as possible to 
  \(\left|\phi\right\rangle \left|\phi\right\rangle\). 
The best cloning machine is one, which maximizes the quantity 
\(F_{clone} = (\left\langle \psi \right| \left\langle \psi \right| \left| \Psi\right\rangle
+ \left\langle \phi \right| \left\langle \phi \right| \left| \Phi\right\rangle)/2\) \cite{Buzek}.
In the case of deleting, we want to have the \emph{complementary} evolution
\(\left|\psi\right\rangle \left|\psi\right\rangle \rightarrow \left|\Psi_d\right\rangle\) 
and \(\left|\phi\right\rangle \left|\phi\right\rangle \rightarrow \left|\Phi_d\right\rangle\)
(in a closed system), where
in the perfect case, we want to have 
\(\left|\Psi_d\right\rangle  =\left|\psi\right\rangle  \left|0\right\rangle\) and  
 \(\left|\Phi_d\right\rangle  = \left|\phi\right\rangle \left|0\right\rangle \),
 \(\left|0\right\rangle\) being a fixed state from which information (whether it was 
\(\left|\psi\right\rangle\) or \(\left|\phi\right\rangle\)) has been deleted.
Again this exact case is not possible under a quantum mechanical operation, when 
\(\left|\psi \right\rangle\) and \(\left|\phi\right\rangle\) are nonorthogonal \cite{Braun_Pati, Braun_Pati2}. 
So just as in the case of cloning, one may again want to obtain 
\(\left|\Psi_d\right\rangle\) 
as close as possible to \(\left|\psi\right\rangle  \left|0\right\rangle\), and  
at the same time 
\(\left|\Phi_d\right\rangle\) as close as possible to 
  \(\left|\phi\right\rangle \left|0\right\rangle\). 
The best deleting machine is one that, for some fixed \(\left|0\right\rangle\),
maximizes the 
quantity 
\(F_{delete} = (\left\langle \psi \right| \left\langle 0 \right| \left| \Psi_d\right\rangle
+ \left\langle \phi \right| \left\langle 0 \right| \left| \Phi_d\right\rangle)/2\) (cf. \cite{Braun_Pati2}).

We now 
show that 
\begin{theorem}
In a physical system that follows (i), (ii), and (iii),
for two nonorthogonal states (\(\left|\psi\right\rangle\) and \(\left|\phi\right\rangle\)), cloning 
evolutions that allow fidelities that are better than 
the best quantum mechanically attainable fidelity \(F_{clone}\), will result in signaling.  
\end{theorem}

Before proving the theorem, let us note that in \cite{Gisin_signaling} (cf. 
\cite{Bruss_signaling}), 
it was shown that a better fidelity than the best quantum mechanical fidelity leads to 
signaling. And Ref. \cite{Braun_Pati3} shows 
that exact deleting results in signaling. However 
in both these cases, they considered \emph{universal} cloning and deleting. Such 
cloning and deleting are invalidated by linearity. Here however we consider 
cloning and deleting of two nonorthogonal states, which cannot be ruled out by linearity. No cloning 
and no deleting of two nonorthogonal states can be proven
by using unitarity, a more stricter restriction than just linearity. It has been 
widely regarded that violation of linearity will lead to signaling  (cf. \cite{thhik-ki-na}). 
Our results show that important linear 
operations can also lead to signaling.


\textbf{Proof.} 
Let us consider symmetric cloning. However all the considerations carry over, with a little
more algebra, 
to the asymmetric case also.
Suppose that for the input states \(\left|\psi\right\rangle\) and \(\left|\phi\right\rangle\),
the best quantum mechanically attainable cloning fidelity is 
\(F_{clone}\), and is attained with the states 
\(\left|\Psi\right\rangle\) and \(\left|\Phi\right\rangle\). 
Suppose also that there exists a 
(non-quantum)
 cloning  machine that produces the states 
\(\left| \Psi{'}\right\rangle\) and \(\left|\Phi{'}\right\rangle\), 
giving
 a better fidelity 
\(F{'}_{clone}= (\left\langle \psi \right| \left\langle \psi \right| \left|\Psi{'}\right\rangle
+ \left\langle \phi \right| \left\langle \phi \right| \left| \Phi{'}\right\rangle)/2\), that is 
\(>F_{clone}\).

\begin{figure}[tbp]
  \epsfig{figure=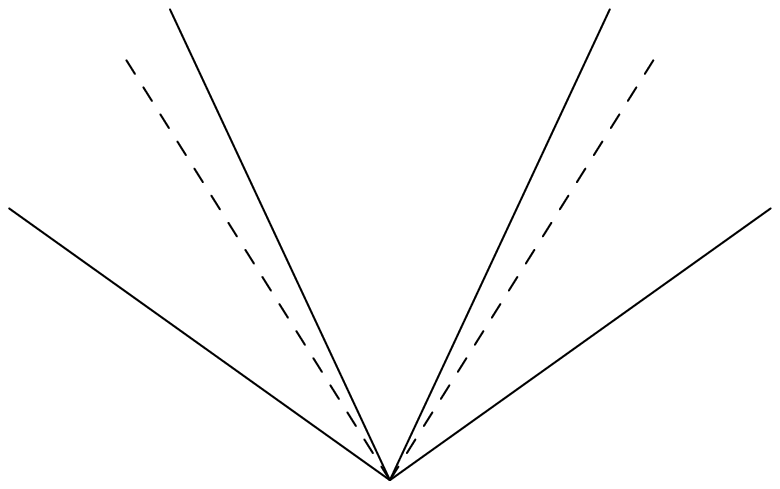,width=0.25\textwidth}
\put(0,46){\(\phi \phi\)}
\put(-140,46){\(\psi \psi\)}
\put(-100,80){\(\Psi \)}
\put(-35,80){\(\Phi \)}
\put(-117,70){\(\Psi{'} \)}
\put(-20,70){\(\Phi{'} \)}
\caption{A pictorial representation of the states \(\left|\psi\right\rangle \left| \psi \right\rangle\),
\(\left|\phi\right\rangle \left| \phi \right\rangle\), \(\left| \Psi \right\rangle\),  
\(\left|\Phi\right\rangle\), \(\left| \Psi{'} \right\rangle\), and 
\(\left|\Phi{'}\right\rangle\). 
} 
\label{cloning_figure}
\end{figure}

In Fig. \ref{cloning_figure}, we give a pictorial representation of the states 
\(\left|\psi\right\rangle \left| \psi \right\rangle\),
\(\left|\phi\right\rangle \left| \phi \right\rangle\), \(\left| \Psi \right\rangle\), 
\(\left|\Phi\right\rangle\), \(\left| \Psi{'} \right\rangle\), and 
\(\left|\Phi{'}\right\rangle\). Note that in general, e.g.
\(\left| \Psi \right\rangle\) and 
\(\left|\Phi\right\rangle\) will not be in the same plane as 
\(\left|\psi\right\rangle \left| \psi \right\rangle\) and
\(\left|\phi\right\rangle \left| \phi \right\rangle\). 
Consider the cone formed by 
\(\left|\Psi\right\rangle\) and 
\(\left|\Phi\right\rangle \). The angle (modulus of inner product) between these states 
must be 
the same as that between 
\(\left|\psi\right\rangle \left| 0 \right\rangle\) and
\(\left|\phi\right\rangle \left| 0 \right\rangle\). This is due to the fact that unitary 
evolution preserves the inner product of evolved states. So 
\(\left|\psi\right\rangle \left| 0 \right\rangle\) and
\(\left|\phi\right\rangle \left| 0 \right\rangle\) must lie on the same cone as that of 
\(\left|\Psi\right\rangle\) and 
\(\left|\Phi\right\rangle \).    
Now whenever \(\left|\psi\right\rangle\) and \(\left|\phi\right\rangle\) are 
nonorthogonal, we have 
\(
\left|\left\langle \psi \right| \left\langle \psi \right| \left| \phi \right\rangle \left| \phi \right\rangle\right|
< \left|\left\langle \psi \right| \left\langle 0 \right| \left| \phi \right\rangle \left| 0\right\rangle\right|
= \left|\left\langle \Psi \right|  \left| \Phi \right\rangle \right|.
\)
This is why the cone of  
\(\left|\psi\right\rangle \left| \psi \right\rangle\) and
\(\left|\phi\right\rangle \left| \phi \right\rangle\) is drawn to be wider than the cone of 
\(\left|\Psi\right\rangle\) and
\(\left|\Phi\right\rangle\) in Fig. \ref{cloning_figure}.

As \(F{'}_{clone} > F_{clone}\), the cone formed by \(\left|\Psi{'}\right\rangle\)
and \(\left|\Phi{'}\right\rangle\) will be wider than that formed by 
\(\left|\Psi\right\rangle \) and
\(\left|\Phi\right\rangle \)
(see Fig. \ref{cloning_figure}). 
Since we consider symmetric cloning, all three cones will be coaxial. 
Thus we have
\(
\left|\left\langle \Psi{'} \right| \left| \Phi{'} \right\rangle \right|
< \left|\left\langle \Psi \right|  \left| \Phi \right\rangle\right|.
\)
But 
\(
\left|\left\langle \Psi \right| \left| \Phi \right\rangle \right|
= \left| \left\langle \psi \right| \left\langle 0 \right|  \left| \phi \right\rangle \left| 0 \right\rangle \right|
\), since 
\(\left| \Psi \right\rangle\) and \( \left| \Phi \right\rangle\) are produced from 
\(\left| \psi \right\rangle \left| 0 \right\rangle\) and 
\( \left| \phi \right\rangle \left| 0 \right\rangle\) by quantum mechanical operations.
Therefore we have that 
\(\left|\left\langle \Psi{'} \right| \left| \Phi{'} \right\rangle \right|
< \left| \left\langle \psi \right| \left\langle 0 \right|  \left| \phi \right\rangle \left| 0 \right\rangle \right|\),
a clear departure from quantum mechanical evolutions (since inner product must be 
preserved in quantum mechanical evolutions). 
And whenever this relation 
holds, the von Neumann entropy of 
\(\varrho_{out} = 
\left(\left|\Psi{'}\right\rangle \left\langle \Psi{'} \right| 
+ \left|\Phi{'}\right\rangle \left\langle \Phi{'} \right| \right)/2\)
is \emph{greater} than the von Neumann entropy of 
\(\varrho_{in} = 
\left(\left|\psi\right\rangle \left|0\right\rangle 
\left\langle \psi \right| \left\langle 0 \right| 
+ \left|\phi\right\rangle \left|0\right\rangle 
\left\langle \phi \right| \left\langle 0 \right|  \right)/2\).

Consider now 
the bipartite state
\begin{equation}
\label{asol}
\left|\alpha\right\rangle = \frac{1}{\sqrt{2}}\left(\left|0\right\rangle_A(\left|\psi\right\rangle
\left|0\right\rangle)_B + \left|1\right\rangle_A(\left|\phi\right\rangle
\left|0\right\rangle)_B
\right),
\end{equation}
where 
\(\langle 0 || 1 \rangle =0\).
Suppose that a super-quantum mechanical cloning evolution, attaining \(F{'}_{clone}\) for 
the states \(\left|\psi\right\rangle\) and \(\left|\phi\right\rangle\),  acts on 
part B 
of the state \(\left|\alpha \right\rangle\),
so that 
 the state \(\left|\alpha \right\rangle\) evolves into 
\(
\left|\alpha_1\right\rangle = 
\left(\left|0\right\rangle_{A}
\left|\Psi{'}\right\rangle_{B}
+ \left|1\right\rangle_{A}
\left|\Phi{'}\right\rangle_{B}
\right)/\sqrt{2} 
\).
Note that we have explicitly used linearity (item (iii)), in obtaining the state \(\left|\alpha_1\right\rangle\).
The local density matrices of the B part of the states \(\left|\alpha\right\rangle\)
and \(\left|\alpha_1\right\rangle\)
are \(\varrho_{in}\) and \(\varrho_{out}\). We therefore have a difference in von Neumann 
entropy of the input and output states in the B part. Since 
 \(\left|\alpha\right\rangle\)
and \(\left|\alpha_1\right\rangle\)
 are pure states, this difference can be exactly verified in the A parts.
Therefore, consequent upon action of any member of the family  of super-quantum cloning evolutions
(the family is generated by 
pairs of nonorthogonal states)
in the 
B part,
 an increase in entropy 
can be observed 
in the A part.
\(\Box\)

Similar reasoning holds for the case of deleting also. Only Fig. \ref{cloning_figure} must be 
replaced by one in which an outer cone is formed by 
\(\left|\Psi_d\right\rangle\) and \(\left|\Phi_d\right\rangle\) and an inner one formed by 
\(\left|\psi\right\rangle\left|0\right\rangle\) and \(\left| \phi \right\rangle \left|0\right\rangle\).
The middle cone will again be formed by 
\(\left|\Psi{'}_d\right\rangle\) and \(\left|\Phi{'}_d\right\rangle\). 
Here \(\left|\Psi_d\right\rangle\) and \(\left|\Phi_d\right\rangle\) will represent the states which are 
obtained from 
\(\left|\psi\right\rangle\left|\psi\right\rangle\) and \(\left| \phi \right\rangle \left| \phi \right\rangle\),
by the best quantum mechanical deleting operation, 
assumed to be 
\(F_{delete}\).
Also 
the shared bipartite
 state that must be considered is \(\left|\alpha{'}\right\rangle = \left(\left|0\right\rangle_A(\left|\psi\right\rangle
\left|\psi\right\rangle)_B + \left|1\right\rangle_A(\left|\phi\right\rangle
\left|\phi\right\rangle)_B
\right)/\sqrt{2}\).
In this case, a super-quantum deleting evolution in the 
B part,
results in a decrease of entropy 
in the 
A part, so that
\begin{theorem}
In a physical system that follows (i), (ii), and (iii),
for two nonorthogonal states (\(\left|\psi\right\rangle\) and \(\left|\phi\right\rangle\)), deleting 
evolutions that allow fidelities that are better than 
the best quantum mechanically attainable fidelity \(F_{delete}\), will result in signaling.  
\end{theorem}


We will now show that it is possible to test the effect, by showing that 
%
the states \(|\alpha\rangle\) and \(|\alpha{'}\rangle\) (used in Theorems 1 and 2 above) 
can be 
prepared with current technology.
Photons are as yet the best candidates for quantum communication. 
We give our strategy in terms of the polarization degree of freedom  of photons.

\emph{The case of cloning.}
In this case, we require to prepare the state \(\left|\alpha\right\rangle\) of Eq. (\ref{asol}). 
Let us write it as 
\(\frac{1}{\sqrt{2}}(\left|0\right\rangle_1 \left|\psi\right\rangle_2 + 
\left|1\right\rangle_1 \left|\phi\right\rangle_2)\left|0\right\rangle_4\), 
where  
the photon 1 
is to go to Alice (A) who is in the accessible part of the experiment. The photons 2 and 4 are to be
sent to the environment, and 
will not be directly observed
(see Fig. \ref{expt_figure}).
For nonorthogonal 
\(\left|\psi\right\rangle\) and
\(\left|\phi\right\rangle\), the first part 
\(
(\left|0\right\rangle \left|\psi\right\rangle + 
\left|1\right\rangle \left|\phi\right\rangle)/\sqrt{2}\)
is a nonmaximally entangled state. It can of course be written in Schmidt decomposition as 
\(a\left|0{'}\right\rangle \left|0{''}\right\rangle + 
b\left|1{'}\right\rangle \left|1{''}\right\rangle\), where \(a\) and \(b\) are positive numbers with 
\(a^2 + b^2 =1\). We choose the local axes such that this nonmaximally entangled state is 
\(a\left|V\right\rangle \left|H\right\rangle + 
b\left|H\right\rangle \left|V\right\rangle = \left|\beta\right\rangle\) (say), where 
\(\left|V\right\rangle\) and \(\left|H\right\rangle\) are respectively the 
vertical and horizontal polarizations of a photon. This can  be prepared by 
spontaneous pulsed parametric down conversion \cite{HM1,AUMapp}.

A schematic description of the arrangement is given in Fig. \ref{expt_figure}.
A pump laser is directed towards a down conversion crystal. There is then a certain probability of 
obtaining the state 
\(\left|\psi^{+}\right\rangle =
(\left|V\right\rangle_1 \left|H\right\rangle_2 + 
\left|H\right\rangle_1 \left|V\right\rangle_2)/\sqrt{2}\) 
in the modes 1 and 2 \cite{Hannover-ey_Kolkattaia}.
Subsequently, local fitering operations are performed to create the nonmaximally 
entangled state 
\(\left|\beta\right\rangle = a\left|V\right\rangle_1 \left|H\right\rangle_2 + 
b\left|H\right\rangle_1 \left|V\right\rangle_2\)
in the modes 1 and 2. (These local filtering operations are not shown in the figure.)
After passing through the crystal, the pulse is reflected back to the crystal by a delay mirror 
(see e.g. \cite{Linares}). 
There is again a certain probability of creation of a pair in the state \(\left|\psi^{+}\right\rangle\)
in the modes 3 and 4. We consider only those cases when \emph{both} the pairs are created. The mode 
3 is detected and acts as a trigger to indicate that a photon is actually present in mode 4. The 
polarization of the photon 
in mode 4 is set to vertical by using a polarizer. So the photon in mode 4 is ultimately in the 
state \(\left|V\right\rangle\), and this acts as our blank state \(\left|0\right\rangle_4\)
in the total state \(\left|\alpha\right\rangle_{124} = \left|\beta\right\rangle_{12}\left|0\right\rangle_{4}\).
The mode 4 and the mode 2 (after being reflected by two mirrors) is 
directed to a half-silvered mirror, so that mode 4 passes through and mode 2 is reflected. The delay 
in the creation of the pair 34 is made such that the photons in modes 2 and 4 reach the half-silvered 
mirror at the same time. Then these two photons are directed to the environment.
The photon in mode 2 runs towards Alice (A), and remains in the accessible part of the experiment. 


Here we are using Type II down conversion \cite{Kwiat}. In Type I case, the 
path degrees of freedom are used for entanglement generation. This is a problem here, as 
we want the B part photons to ultimately be directed towards a single direction. 
Note here that we have \emph{not} used entanglement swapping \cite{Marek1,Sougato1} 
to prepare our  entangled state. 
Here, the photon 
3 acts as a trigger for guaranteeing the existence of photon 4, while the photon 1 will 
subsequently be detected by 
Alice (and will act as a trigger for the state created in 
modes 12), and we consider only those runs of the experiment, in  
which both the trigger photon 3 and the photon 1 are detected.

\emph{The case of deleting.} 
In this case, we must prepare the state 
\(\left|\alpha{'}\right\rangle\).
This can be obtained after local filtering operations on a GHZ state \cite{GHZ_original} 
\(
\left(\left|0\right\rangle_A(\left|0\right\rangle
\left|0\right\rangle)_B + \left|1\right\rangle_A(\left|1\right\rangle
\left|1\right\rangle)_B
\right)/\sqrt{2}\), 
after which the first part remains in the accessible part (A) of the experiment and 
the second and third parts are aligned to a single direction (just as in Fig. \ref{expt_figure} 
in
case of cloning) and sent to the environment. Experimental 
observation of the GHZ state has been reported
 in  \cite{GHZ_expt}. However the experiment relies for its success on 
actual observation of \emph{all} the photons that make up the GHZ state (plus
a trigger photon). Whereas this is sufficient for many important purposes, it is not 
sufficient for us. In our case, at least two photons are 
not
to be directly observed. 
However in a proposal for preparation for the GHZ state 
\cite{GHZ_proposal}, the state is prepared without the restriction of having to 
actually detect the photons (making up the GHZ), to know that a GHZ state is produced. 
After production of a GHZ by this proposal, local filtering operations can be 
carried out to produce the state \(\left|\alpha{'}\right\rangle\).

After the photons in the B part are sent to the environment, 
Alice makes measurements on her photon 
to determine the von Neumann entropy of her state. 
The von Neumann entropy can conveniently be found by measurement results from outcomes in 
a Mach-Zehnder interferometer, to which the photon in mode 1 can be directed into. 
More economical methods , although requiring measurements over many copies, can be found in Refs.
\cite{Pawel}.
The von Neumann entropy of the A part of the state 
\(\left|\alpha\right\rangle\),
or the 1 part of the state 
\(\left|\beta \right\rangle_{12} \left| V \right\rangle_4\) is 
\(
H(a^2) = -a^2 \log_2 a^2 - b^2 \log_2 b^2.
\)
Similarly, let the von Neumann entropy of the A part of the state \(\left|\alpha{'}\right\rangle\)
be \(H({a{'}}^2)\).
As we have seen in Theorem 1 above, 
any departure from the value \(H(a^2)\) in the experiment for 
cloning, or from the value \(H({a{'}}^2)\) in the experiment for 
deleting, of the von Neumann entropy of the polarization degrees of freedom of the 
photon 1, as detected by Alice  from her experimental 
results, will indicate a signaling. This in turn indicates that there are 
non-quantum mechanical operations that have acted on the modes 2 and 4, that were 
directed to the environment.

The same experiment can be carried on for different values of \(a\), \(a{'}\). The values of 
\(a\), \(a{'}\)
 can be varied by varying the parameters of the local filtering apparatus. Each set  of \(\{a,a{'}\}\), 
checks for a duo of non-quantum mechanical evolutions, one from 
super-quantum mechanical cloning, and the other from super-quantum mechanical deleting.
Thus we can 
check for two \emph{complementary} families of possible non-quantum mechanical evolutions on the modes 2 and 4. 
In an actual experiment, there will be some noise. The results 
obtained from such experiments can be used to put bounds on the power of possible
non-quantum mechanical evolutions in 
the environment.


In principle, the ``environment'' can be some extreme situations, e.g. an evaporating 
black hole, where conditions may be far too extreme for the laws established in the usual 
laboratories to be applicable (see e.g. \cite{Yurt}, cf. \cite{jothhesthho}). However just as in 
the recent proposal \cite{Yurt}, the way to send the probes (the photons 2 and 4 in our case)
to an evaporating black hole, remains a problem. 
But let us mention that we consider a bipartite state instead of the three-party state 
of \cite{Yurt}. 
Another conceivable situation is where a person claims to be able to perform non-quantum 
operations, but denies direct access to his/her laboratory. Our procedure can then 
be used to check for his/her claim. However the main impetus, 
in this paper
(of the Theorems 1 and 2), is to have an independent reason for believing in the 
quantum dynamics.

In conclusion, we have shown that in a physical system that follows (i), (ii), and (iii), a cloning operation 
acting in a region B, that 
leads to a better than quantum mechanical fidelity, results in signaling to a far-apart region
A.  The same conclusion can be obtained for deleting. The strategy to check for such signaling does not 
require to perform measurements in the region B. The two-party states required to perform the strategy 
can be prepared with current technology. This gives us an independent basis to 
believe in the quantum dynamics.

We acknowledge support from  
the Alexander von Humboldt Foundation, and the
EC Contract 
QUPRODIS.


\begin{thebibliography}{10}
\bibitem{von} The von Neumann entropy of  \(\varrho\) is 
\(S(\varrho)=-\mbox{tr}\varrho \log_2 \varrho\).
\bibitem{Hannover-ey_Kolkattaia} One has to go through the usual quantum evolution to derive
the output state of a parametric down conversion \cite{HM1,AUMapp}. However the form of the output 
state can also be checked in an experiment, without directly going to the quantum formalism. Also, the 
state is produced in the accessible part of our experiment, where we may believe in the quantum formalism, without
necessarily believing it to be true in the environment.
\end{thebibliography}
\end{document}